\documentclass[aps,prc,preprint,tightenlines,showpacs,byrevtex,endfloats,%
nofootinbib]{revtex4}

\usepackage{epsfig,bm,dcolumn}

\begin{document}

\preprint{nucl-th/0204035}


\title{
One-loop corrections to  $\bm{\omega}$ photoproduction
near threshold}


\author{Yongseok Oh}%
\email{yoh@phya.yonsei.ac.kr}

\affiliation{Institute of Physics and Applied Physics,
Department of Physics, Yonsei University, Seoul 120-749, Korea}

\author{T.-S. H. Lee}%
\email{lee@theory.phy.anl.gov}

\affiliation{
Physics Division, Argonne National Laboratory, Argonne,
Illinois 60439}



\begin{abstract}

One-loop corrections to $\omega$ photoproduction near threshold
have been investigated by using the approximation
that all relevant transition amplitudes
are calculated from the tree diagrams of effective Lagrangians.
With the parameters constrained by the data
of $\gamma N \to \pi N$, $\gamma N \to \rho N$, and $\pi N \to \omega N$
reactions, it is found that the one-loop effects due to the intermediate
$\pi N$ and $\rho N$ states can significantly change the differential cross
sections and spin observables.
The results from this exploratory investigation suggest
strongly that the coupled-channel effects should be taken into account in
extracting reliable resonance parameters from the data of vector meson
photoproduction in the resonance region. 

\end{abstract}


\maketitle

\section{Introduction}

Experimental data of vector meson photoproductions are now being rapidly
accumulated at Bonn \cite{Klein96-98}, Thomas Jefferson National Accelerator
Facility \cite{CLAS}, GRAAL of Grenoble \cite{GRAAL01}, and LEPS of
SPring-8 \cite{LEPS01}.  
The study of photoproduction of vector mesons ($\omega, \rho, \phi$) is
expected to be useful to resolve the so-called ``missing resonances''
problem \cite{CR00}.
In addition, the extracted resonance parameters can shed lights on the
structure of nucleon resonances ($N^*$) and can be used to test the
existing hadron models.
In recent years, some theoretical progress has been made
\cite{ZLB98c-Zhao01,OTL01} in this direction.
In this work we will address the question about how these earlier models
should be improved for a more reliable extraction of the $N^*$
parameters from the forthcoming data.
To be specific, we consider the photoproduction of $\omega$ meson.

It is well known that the extraction of $N^*$ parameters from
experimental data depends strongly on the accuracy of the treatment
of the non-resonant amplitudes.
In all of the recent studies of $\omega$ photoproduction at resonance
region \cite{ZLB98c-Zhao01,OTL01}, the non-resonant amplitudes are
calculated from the tree diagrams of effective Lagrangians.
This is obviously not satisfactory for the following reasons. 
First, the tree-diagram models do not include the hadronic final state
interaction (FSI).
The importance of FSI in interpreting the data has been
demonstrated in the study of pion photoproductions.
For example, the calculations of Ref. \cite{SL96} have shown that the
magnetic $M1$ amplitude of the $\gamma N \rightarrow \Delta(1232)$
transition can be identified with the predictions from constituent quark
models only when the pion re-scattering effects (i.e., pion cloud effects),
as required by the unitary condition, are accounted for appropriately 
in analyzing the data.
Second, the vector meson productions occur in the energy region where
several meson-nucleon channels are open and their influence must be
accounted for.
This coupled-channel effect was already noticed and explored in 1970's
for vector meson photoproduction \cite{SS68}.
In this paper we make a first attempt to re-investigate this problem in
conjunction with the approach developed in Ref. \cite{OTL01}. 

In a dynamical formulation, such as that developed in Ref. \cite{SL96},
the most ideal approach is to carry out a coupled-channel calculation.
At energies near the $\omega$ photoproduction threshold, the meson-baryon
channels which must be included in a coupled-channel calculation are many,
such as $\pi N$, $\pi \Delta$, $\rho N$, and $K Y$.
Such a full coupled-channel calculation is not feasible at the
present stage, mainly because some of the experimental information 
that are needed to constrain the transitions between relevant hadronic
meson-baryon channels are not available.
For example, there is no information about $\pi \Delta \to \omega N$ and
$K Y \to \omega N$ transitions.
We therefore are only able to consider just the effects due to
intermediate $\pi N$ and $\rho N$ channels. 
In this exploratory investigation, we will follow Ref. \cite{SS68}
to further simplify the calculations
by  only considering the one-loop corrections which are the leading order
terms in a perturbation expansion of a full coupled-channel formulation,
as will be explained in Sec. II.
Nevertheless, our results will shed some lights on the
importance of coupled-channel effects and provide information
for developing a much more complex full coupled-channel calculation.
In many respects, our investigation is similar to a recent investigation
of coupled-channel effects on kaon photoproduction \cite{CTLS01}.

This paper is organized as follows.
In Sec. II, we introduce a coupled-channel formulation of
$\gamma N$ reaction and indicate the procedures for calculating
the one-loop corrections to  $\omega$ photoproduction.
Section III is devoted to specify various transition amplitudes
which will be used as the inputs to our calculations. 
Numerical results are presented and discussed in Sec. IV. 
The conclusions are given in Sec. V.
Some details on the $\pi N \to \omega N$ reaction are given in Appendix.

\section{Dynamical Coupled-Channel Formulation}

In the considered energy region, the $\gamma N$ reaction is a
multi-channel multi-resonance problem.
In this work we follow the dynamical approach developed by Sato and Lee
\cite{SL96} to investigate this problem.
It is done by simply extending the scattering formulation of Ref. \cite{SL96}
to include more $N^*$ states and more meson-nucleon channels.
The resulting amplitude $T_{\gamma N,\omega N}(E)$ for the
$\gamma N \rightarrow \omega N$ reaction can be written as 
\begin{eqnarray}
T_{\gamma N,\omega N}(E)  =  t_{\gamma N,\omega N}(E) + \sum_{N^*}
\frac{ \bar{\Gamma}_{\gamma N \rightarrow N^*}
\bar{\Gamma}_{N^* \rightarrow \omega N} }
{E -  M_{N^*}^0  - \Sigma_{N^*}(E)},
\label{Ttotal}
\end{eqnarray}
where $t_{\gamma N,\omega N}$ is the non-resonant amplitude. 
It is defined by the following coupled-channel equations
\begin{eqnarray}
t_{\gamma N, \omega N}&= &B_{\gamma N, \omega N}
 +\sum_{\alpha} B_{\gamma N,\alpha}G_{\alpha}(E)
 t_{\alpha,\omega N},
\label{tGN} \\ 
t_{\alpha,\beta}&= &v_{\alpha,\beta} 
+ \sum_{\delta} v_{\alpha,\delta}G_{\delta}(E) 
t_{\delta,\beta},
\label{tAB}
\end{eqnarray}
where $\alpha,\beta$ denote the considered meson-nucleon channels
such as $\omega N$, $\pi N$, $\rho N$, $\pi \Delta$, and $KY$.
$B_{\gamma N,\alpha}$ is the non-resonant photoproduction amplitude,
$v_{\alpha,\beta}$ are the non-resonant meson-nucleon interactions, and
$G_\alpha$ is the free meson-nucleon propagator defined by 
\begin{eqnarray}
G_{\alpha}(E)=\frac{1}{E - \left( H_0 \right)_{\alpha} + i\epsilon}.
\label{Green}
\end{eqnarray}
Here $(H_0)_\alpha$ is the free Hamiltonian in channel $\alpha$.
For channels containing an unstable particle, such as $\rho N$ and
$\pi \Delta$, their widths must be included appropriately. Here we follow the
procedure of Ref. \cite{LEE84}.

The $N^*$ excitations are described by the second term of Eq.
(\ref{Ttotal}).
It is defined by the dressed vertex functions 
\begin{eqnarray}
\bar{\Gamma}_{\gamma N \rightarrow N^*}  &=&  
  { \Gamma_{\gamma N \rightarrow N^*}} + \sum_{\alpha}
v_{\gamma N,\alpha}
G_{\alpha}(E)\bar{\Gamma}_{\alpha \rightarrow N^*},
\nonumber \\
\bar{\Gamma}_{N^* \rightarrow \omega N}
 &=&  \Gamma_{N^* \rightarrow \omega N} +
\sum_{\alpha} \Gamma_{N^*\rightarrow \alpha}
G_{\alpha}(E)t_{\alpha,\omega N},
\label{vertex}
\end{eqnarray}
and the $N^*$ self-energy
\begin{eqnarray}
\Sigma_{N^*}(E) &=& \sum_{\alpha}
\bar{\Gamma}_{N^*\rightarrow \alpha}
G_{\alpha}(E) \Gamma_{\alpha \rightarrow N^*}.
\end{eqnarray}
The bare mass $M^0_{N^*}$ of Eq. (\ref{Ttotal}) and the bare vertices
$\Gamma_{\gamma N \leftrightarrow N^*}$ and
$\Gamma_{\alpha \leftrightarrow N^*}$ of Eq. (\ref{vertex}) can be
identified with the predictions from a hadron
model that does not include the continuum meson-baryon states.

In this paper, we focus on the calculation of the non-resonant amplitudes
defined by Eqs.~(\ref{tGN}) and (\ref{tAB}).
The extraction of resonance parameters from the data depends heavily on
the accuracy of this dynamical input.
At energies near the $\omega$ production threshold, the hadronic
meson-baryon channels that must be included in solving
Eqs. (\ref{tGN}) and (\ref{tAB}) are many, such as $\pi N$, $\omega N$,
$\rho N$, $\pi \Delta$, and $K Y$. 
Because of the data which are needed to constraint the interaction
$v_{\alpha,\beta}$ of Eq. (\ref{tAB}) are very limited,
we are only able to consider the effects due
to the intermediate $\pi N$ and $\rho N$ channels.
To further simplify the investigation, we make the one-loop approximation that
the amplitude in Eq. (\ref{tGN}) is evaluated by setting 
$t_{\alpha,\beta} \sim v_{\alpha,\beta}$.
No attempt is made to solve the coupled-channel equation (\ref{tAB}).
Furthermore, we assume that the interaction $v_{\alpha,\beta}$ can be
calculated from the tree diagrams of effective Lagrangians.
This is certainly not very satisfactory, but it should be sufficient for
this exploratory study.


\begin{figure}
\centering
\epsfig{file=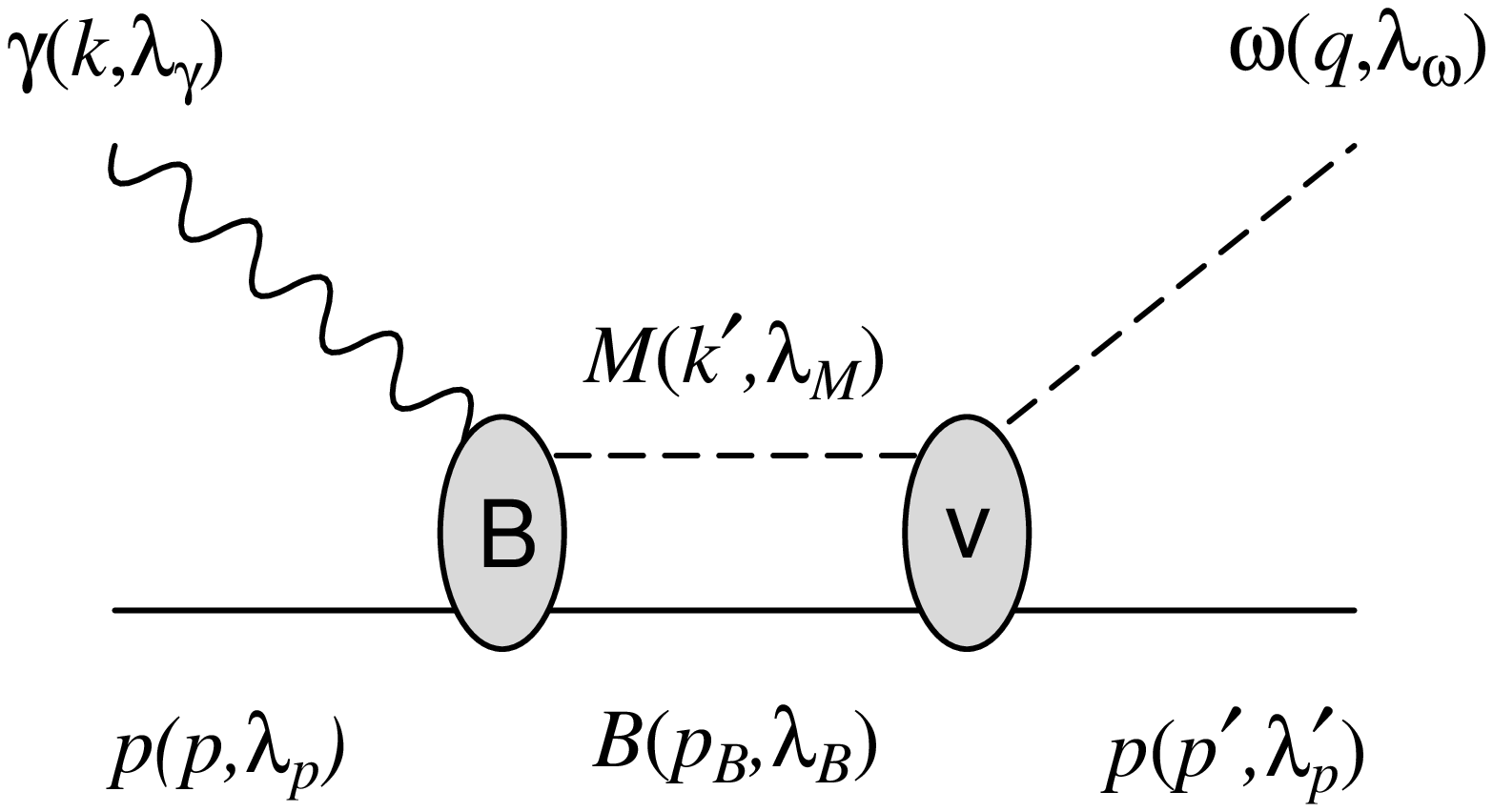, width=8cm}
\caption{Diagramatic representation of the intermediate meson-baryon (MB)
state in $\omega$ photoproduction.}
\label{fig:FSI}
\end{figure}

Our task in this work is therefore to investigate the non-resonant amplitude
$t_{\gamma N, \omega N}$ defined by Eq. (\ref{tGN}), with
$t_{\alpha,\omega N}$ replaced by $v_{\alpha,\omega N}$.
The second term of Eq. (\ref{tGN}) is then the one-loop correction represented 
graphically in Fig.~\ref{fig:FSI}. 
Explicitly, the matrix element of
this one-loop amplitude in the center of mass frame is
\begin{eqnarray}
t^{\mbox{\scriptsize one-loop}}_{\gamma N,\omega N}({\bf k},{\bf q};E)=
\sum_{M=\pi,\rho}\int d{\bf q^\prime}
B_{\gamma N,M N}({\bf k},{\bf q^\prime};E)G_{MN}({\bf q^\prime},E)
v_{M N,\omega N}^{}({\bf q^\prime},{\bf q};E),
\label{loop-int}
\end{eqnarray}
where ${\bf k}$ and ${\bf q}$ are the momenta for the incoming photon and
the intermediate mesons, respectively.
The  propagator for the $\pi N$ state is 
\begin{eqnarray}
G_{\pi N}({\bf q'},E)=
\frac{1}{E- E_{N}({\bf q^\prime})-E_\pi({\bf q^\prime})+i\epsilon}.
\end{eqnarray}
For the $\rho N$ propagator, we account for the width of the $\rho$ by using
the the approach of Ref. \cite{LEE84}. Neglecting the energy-dependence
of the mass shift term, the $\rho N$ propagator takes the following form
\begin{eqnarray}
G_{\rho N}({\bf q'},E)=
\frac{1}{E- E_{N}({\bf q^\prime})-E_\rho({\bf q^\prime})
+i\frac{\Gamma\bm{(}\omega(q',E)\bm{)}}{2}\theta[\omega^2(q',E)-4M^2_\pi]},
\end{eqnarray}
where $\omega(q',E)=[(E-E_N(q'))^2-q^{'2}]^{1/2}$ is the energy available for 
the $\rho$ meson in its rest frame, and
the step function is $\theta(x)=1$ for $x \geq 0$ and $0$ otherwise.
The width is
\begin{eqnarray}
\Gamma(\omega)=\Gamma^0_\rho\frac{k^3E_\pi(k)}{k^3_0E_\pi(k_0)}
\left(\frac{\Lambda^2_\rho+k^2_0}{\Lambda^2_\rho+k^2}\right)^4,
\end{eqnarray}
where $k$ is defined by $\omega = 2 E_\pi(k)$ and $k_0$ by
 $M_\rho = 2 E_\pi(k_0)$. 
(The above form can be derived from a resonant model for fitting the $\pi\pi$
scattering phase shifts in the  $J=I=1$ channel with a $\rho \to \pi\pi$
vertex interaction.)
We set $\Gamma^0_\rho= 150$ MeV and $\Lambda_\rho= 0.5 $ GeV.

In the following Section, we describe how the matrix elements of
$B_{\gamma N,MN}$ and $v_{MN,\omega N}^{}$ are calculated from effective
Lagrangians and constrained by experimental data.

\section{Non-Resonant Amplitudes}

We assume that all non-resonant amplitudes $B_{\gamma N, MN}$ and
$v_{MN,\omega N}^{}$ in the one-loop term in Eq. (\ref{loop-int}) can be
calculated from the tree-diagrams defined by the following effective
Lagrangian,
\begin{equation}
{\cal L} = {\cal L}_{V\gamma\varphi} + {\cal L}_{VV\varphi} + {\cal
L}_{\varphi NN} + {\cal L}_\sigma + {\cal L}_{\gamma NN} + {\cal
L}_{VNN},
\label{lagrangian}
\end{equation}
where
\begin{eqnarray}
{\cal L}_{V\gamma\varphi} &=& \frac{eg_{\rho\gamma\pi}}{2M_\rho}
\varepsilon^{\mu\nu\alpha\beta} \,\mbox{Tr}\,[\partial_\mu \rho_\nu
\partial_\alpha A_\beta \pi] +
\frac{eg_{\omega\gamma\pi}}{2M_\omega}
\varepsilon^{\mu\nu\alpha\beta} \,\mbox{Tr}\,[\partial_\mu \omega_\nu
\partial_\alpha A_\beta \pi \tau^3]
\nonumber \\ && \mbox{}  +
\frac{eg_{\omega\gamma\eta}}{M_\omega}
\varepsilon^{\mu\nu\alpha\beta} \partial_\mu \omega_\nu \partial_\alpha
A_\beta \eta,
\nonumber \\
{\cal L}_{VV\varphi} &=& \frac{g_{\omega\rho\pi}}{2}
\varepsilon^{\mu\nu\alpha\beta} \,\mbox{Tr}\, [\partial_\mu \omega_\nu 
\partial_\alpha \rho_\beta \pi ],
\nonumber \\
{\cal L}_{\varphi NN} &=& \frac{g_{\pi NN}}{2M_N} \bar\psi \gamma^\mu
\gamma_5 \partial_\mu \pi \psi + 
\frac{g_{\eta NN}}{2M_N} \bar\psi \gamma^\mu \gamma_5 \psi \partial_\mu
\eta,
\nonumber \\
{\cal L}_\sigma &=& g_{\sigma NN} \bar\psi \sigma \psi +
\frac{eg_{\rho\gamma\sigma}}{2M_\rho} \,\mbox{Tr}\, [
\tau^3 \partial_\mu \rho_\nu \left(
\partial^\mu A^\nu - \partial^\nu A^\mu \right) \sigma],
\nonumber \\
{\cal L}_{\gamma NN} &=& e \bar\psi \left( \gamma_\mu \frac{1+\tau_3}{2}
A^\mu - \frac{\kappa_N}{2M_N} \sigma^{\mu\nu} \partial_\nu A_\mu \right)
\psi,
\nonumber \\
{\cal L}_{VNN} &=& \frac{g_{\rho NN}}{2} \bar\psi \left( \gamma_\mu
\rho^\mu - \frac{\kappa_\rho}{2M_N} \sigma^{\mu\nu} \partial_\nu
 \rho_\mu \right) \psi
+ g_{\omega NN} \bar\psi \left( \gamma_\mu
\omega^\mu - \frac{\kappa_\omega}{2M_N} \sigma^{\mu\nu} \partial_\nu
 \omega_\mu \right) \psi,
\label{lags}
\end{eqnarray}
where $\pi$ ($ = \bm{\tau}\cdot \bm{\pi}$), $\eta$, $\rho_\mu$
($ = \bm{\tau} \cdot \bm{\rho}_\mu$), and $\omega_\mu$ are the pion,
eta, rho, and omega meson fields, respectively.
The photon field is represented by $A_\mu$, and $\psi$ and $\sigma$ are
the nucleon and $\sigma$ meson fields, respectively.
$M_N$ ($M_V$) is the nucleon (vector meson) mass and $\kappa_N$ is the
anomalous magnetic moment of the nucleon,
$\kappa_p = 1.79$ and $\kappa_n = -1.91$.
Throughout this work we use the convention that $\varepsilon^{0123} = +1$.

The coupling constants of the Lagrangian (\ref{lags}) are determined
as follows.
First the coupling constants in ${\cal L}_{V\gamma\varphi}$ are determined
by the vector meson radiative decay widths
given by the Particle Data Group (PDG) \cite{PDG00}.
In $\mathcal{L}_{VV\varphi}$, the coupling $g_{\omega\rho\pi}$ has been
estimated by many models including the massive Yang-Mills approach
\cite{KRS84-JJMP88}, the hidden gauge approach \cite{FKTU85}, 
the vector meson dominance model \cite{KKW96}, the unitary effective
resonance model \cite{KVR01}, and QCD sum rules \cite{EIK83}.
All of these models predict that the value of $g_{\omega\rho\pi}$ is in
the range of $10$--$16$ GeV$^{-1}$.
In this work we use $g_{\omega\rho\pi} = 12.9$ GeV$^{-1}$ \cite{TKR01}.

For $\mathcal{L}_{\varphi NN}$, we use the well-known value
$g_{\pi NN}^2/4\pi = 14.0$ and $g_{\eta NN}^2/4\pi = 0.99$ determined
\cite{OTL01,TLTS99} by using the SU(3) relation.
For $\sigma$ meson, its mass and couplings to the nucleon and the vector
mesons are highly model-dependent.
Following Ref. \cite{FS96}, we set $M_\sigma = 0.5$ GeV and determine the
coupling constants of the $\sigma$ meson by reproducing the experimental
data of $\rho$ photoproduction at low energies, as explained in
Refs. \cite{FS96,OTL00}.
Since the branching ratio of $\omega \to \pi\pi\gamma$ is very small
\cite{PDG00}, we do not consider the $\omega\gamma\sigma$ coupling in
this model.

Following Refs. \cite{OTL01,OTL00}, the values of the coupling constants
$g_{\omega NN}$ and $g_{\rho NN}$ are taken from the analyses of $\pi N$
scattering, pion photoproduction, and nucleon-nucleon scattering
\cite{SL96,RSY99}.
All of the coupling constants used in our calculations are summarized
in Table~\ref{tab:cc}.


\begin{table}
\caption{\label{tab:cc} Coupling constants of the effective Lagrangian
(\ref{lagrangian}).}
\begin{center}
\begin{ruledtabular}
\begin{tabular}{c|c||c|c||c|c}
coupling & value & coupling & value & coupling & value \\ \hline
$g_{\rho\gamma\pi}$ & 0.70 & $g_{\pi NN}$ & 13.26 & $g_{\rho NN}$ &
6.12 \\
$g_{\omega\gamma\pi}$ & 1.82 & $g_{\eta NN}$ & 3.53 & $\kappa_\rho$ &
3.1 \\
$g_{\omega\gamma\eta}$ & 0.42 & $g_{\sigma NN}$ & 10.03 & $g_{\omega NN}$ &
10.35 \\
$g_{\omega\rho\pi}$ & 12.9\footnote{in GeV$^{-1}$ unit} &
$g_{\rho\gamma\sigma}$ & 3.0  &
$\kappa_\omega$ & 0.0 \\
\end{tabular}
\end{ruledtabular}
\end{center}
\end{table}

In addition to the tree-diagrams which can be calculated by using the
Lagrangian (\ref{lags}), we also include the Pomeron exchange in the
amplitudes of vector meson photoproduction \cite{DL84-92,LM95,PL97},
although its contribution is relatively small at low energies.
The details of the Pomeron exchange can be found, for example, in
Refs. \cite{TOYM98,OTL01}, and will not be repeated here.


\begin{figure}
\centering
\epsfig{file=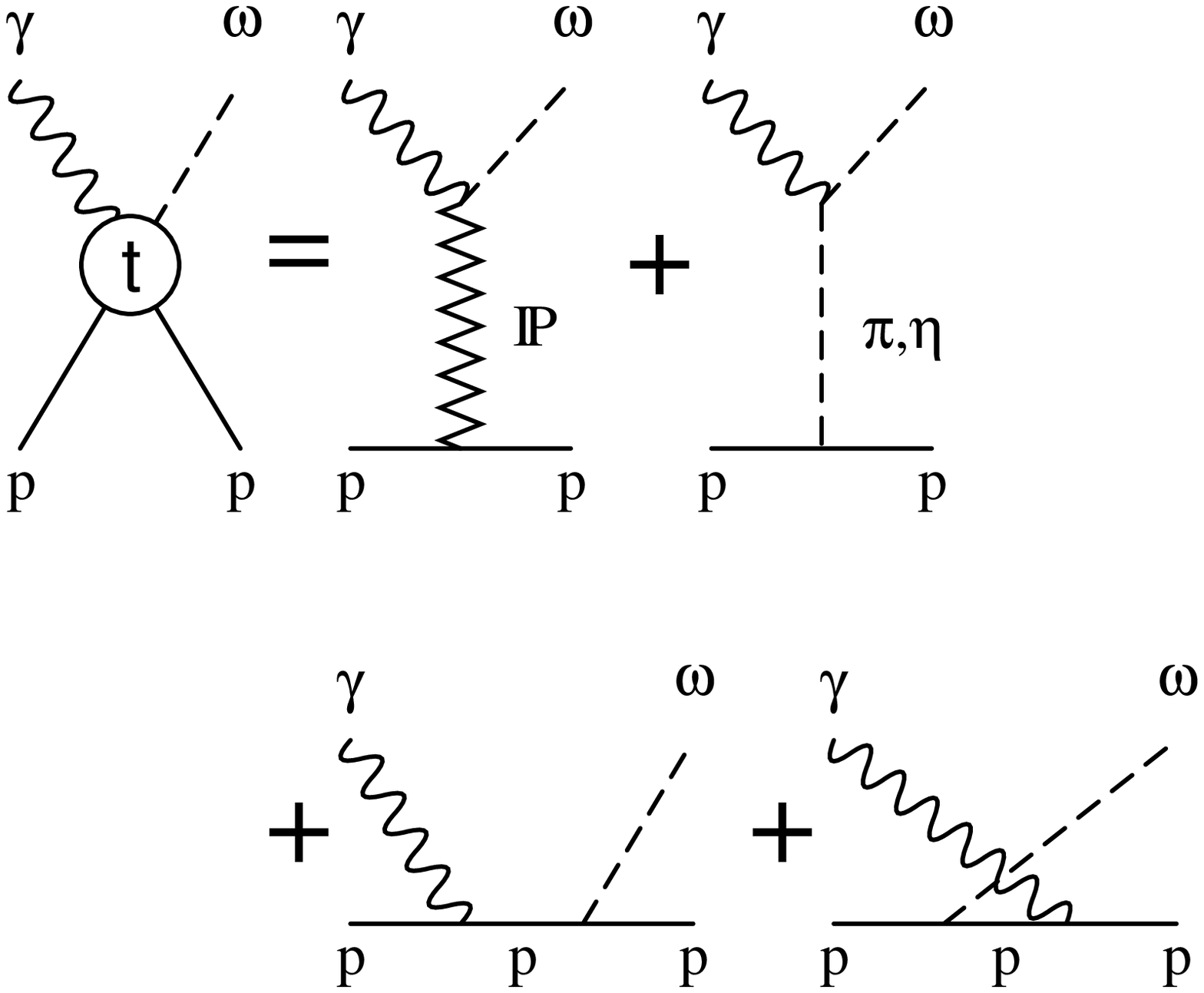, width=8cm}
\caption{
Tree-diagrams for $\gamma p \to \omega p$ which include Pomeron exchange,
$\pi$ and $\eta$ exchange, and the direct and crossed nucleon terms.}
\label{fig:gam-om}
\end{figure}


\begin{figure}
\centering
\epsfig{file=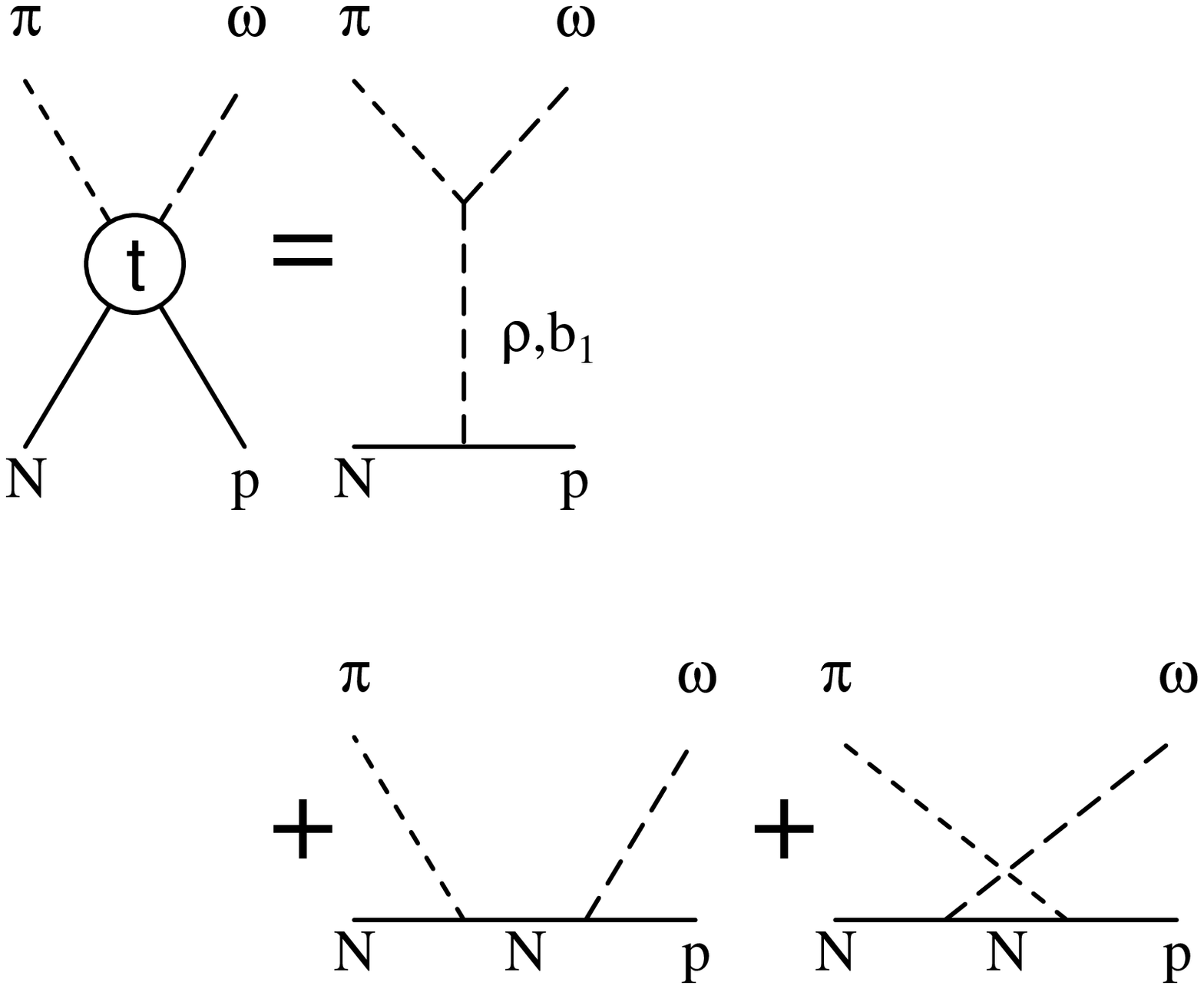, width=8cm}
\caption{Tree-diagrams for $\pi N \to \omega N$ which include $\rho$
and $b_1$ exchanges and the nucleon pole terms.}
\label{fig:pi-omega}
\end{figure}


\begin{figure}
\centering
\epsfig{file=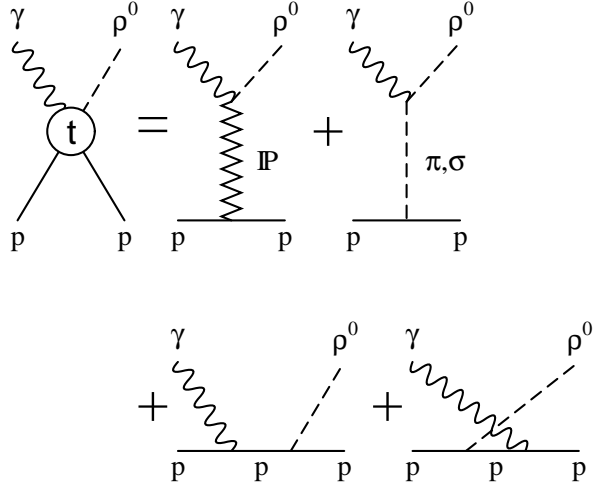, width=8cm}
\caption{Tree-diagrams for $\gamma p \to \rho p$ which include Pomeron,
$\pi$, and $\sigma$ meson exchanges and the nucleon pole terms.}
\label{fig:gam-rho}
\end{figure}


\begin{figure}
\centering
\epsfig{file=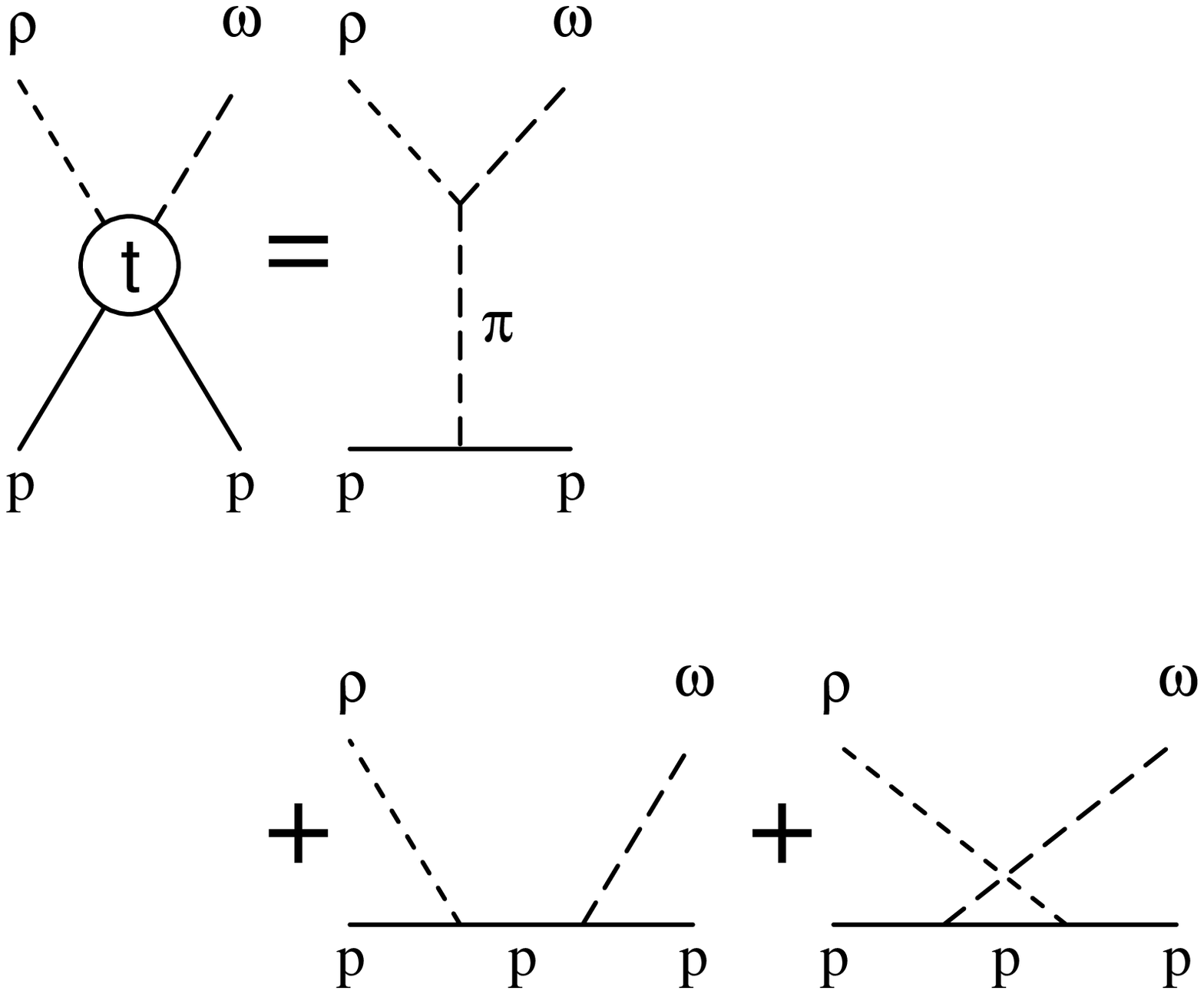, width=8cm}
\caption{Tree-diagrams for $\rho p \to \omega p$ which include one-pion
exchange and the nucleon pole terms.}
\label{fig:rho-omega}
\end{figure}

The considered tree-diagrams are then illustrated
in Fig.~\ref{fig:gam-om} for $\gamma p \rightarrow \omega p$,
Fig.~\ref{fig:pi-omega} for $\pi N \rightarrow \omega p$,
Fig.~\ref{fig:gam-rho} for $\gamma p \rightarrow \rho^0 p$,
and Fig.~\ref{fig:rho-omega} for $\rho^0 p \rightarrow \omega p$.
The calculations of these tree-diagrams are straightforward and therefore
are not detailed here.
These amplitudes are regularized by form factors as follows.
For the $t$-channel exchanges, we use
\begin{equation}
F_t(t) = \frac{\Lambda^2 - M_{\rm ex}^2}{\Lambda^2 - t}
\label{Ft}
\end{equation}
for each vertex, where $M_{\rm ex}$ is the mass of the exchanged particle.
For the $s$ and $u$ channel diagrams, we include \cite{PJ91}
\begin{equation}
F_{su} (r) = \frac{\Lambda_B^4}{\Lambda_B^4 + (r-M_B^2)^2},
\end{equation}
where $r = (s,u)$ and $M_B$ is the mass of the intermediate baryon,
i.e., the nucleon in our case.
For the cutoff parameters in the tree-diagrams for $\gamma p \to \omega
p$, we use the values adopted in Ref. \cite{OTL01},
\begin{equation}
\Lambda_{\pi NN} = 0.6, \qquad
\Lambda_{\eta NN} = 1.0, \qquad
\Lambda_{\omega\pi\gamma} = 0.77, \qquad
\Lambda_{\omega\eta\gamma} = 0.9, \qquad
\Lambda_N = 0.5
\label{cutoff:gam-om}
\end{equation}
in GeV unit and the other cutoff parameters used for the other reactions
will be specified later in Sec.~\ref{sec:res}.
The gauge invariance of the nucleon pole terms are restored by making
use of the projection operators as in Ref. \cite{OTL01}.

The $\gamma N \to \pi N$ amplitudes are not discussed here because 
no tree-diagram model until now can describe the data in the considered
energy region.
Instead we construct the non-resonant amplitudes for $\gamma p \to \pi^0p$
and $\gamma p \to \pi^+ n$ by subtracting the resonance amplitudes
from the empirical multipole amplitudes of the SAID program \cite{said}.
The $\gamma N \rightarrow \pi N$ resonant amplitudes are calculated by
using the procedure given in Ref. \cite{DGL02} except that we use the
resonance parameters from PDG, not from those of Capstick and Roberts
\cite{Caps92-CR94}.
Clearly, this is very model-dependent approach, but should be sufficient
for this very exploratory investigation.
The procedures introduced above only define the on-shell matrix elements
of $\gamma N \rightarrow \pi N$ transition.
For the loop integration (\ref{loop-int}) we need to define its
off-shell behavior.
Guided by the work of Ref. \cite{SL96}, we assume that
\begin{eqnarray}
B_{\gamma N, \pi N}({\bf k}, {\bf q})= B_{\gamma N, \pi N}({\bf k},{\bf q_0})
\left(\frac{\Lambda^2+q^2_0}{\Lambda^2+q^2}\right)^2,
\label{offshell}
\end{eqnarray}
where $\Lambda=0.5 $ GeV is chosen.
To be consistent, the off-shell extrapolation Eq. (\ref{offshell}) is also
used in the loop integration over $\rho N$ state.

\section{\label{sec:res} Results and Discussions}

We can now perform the calculations based on Eqs. (\ref{Ttotal}) and
(\ref{loop-int}).
To proceed, the resonant term of Eq. (\ref{Ttotal}) can be fixed by 
using the quark model predictions \cite{Caps92-CR94} and the
procedures detailed in Ref. \cite{OTL01}.

\begin{figure}
\centering
\epsfig{file=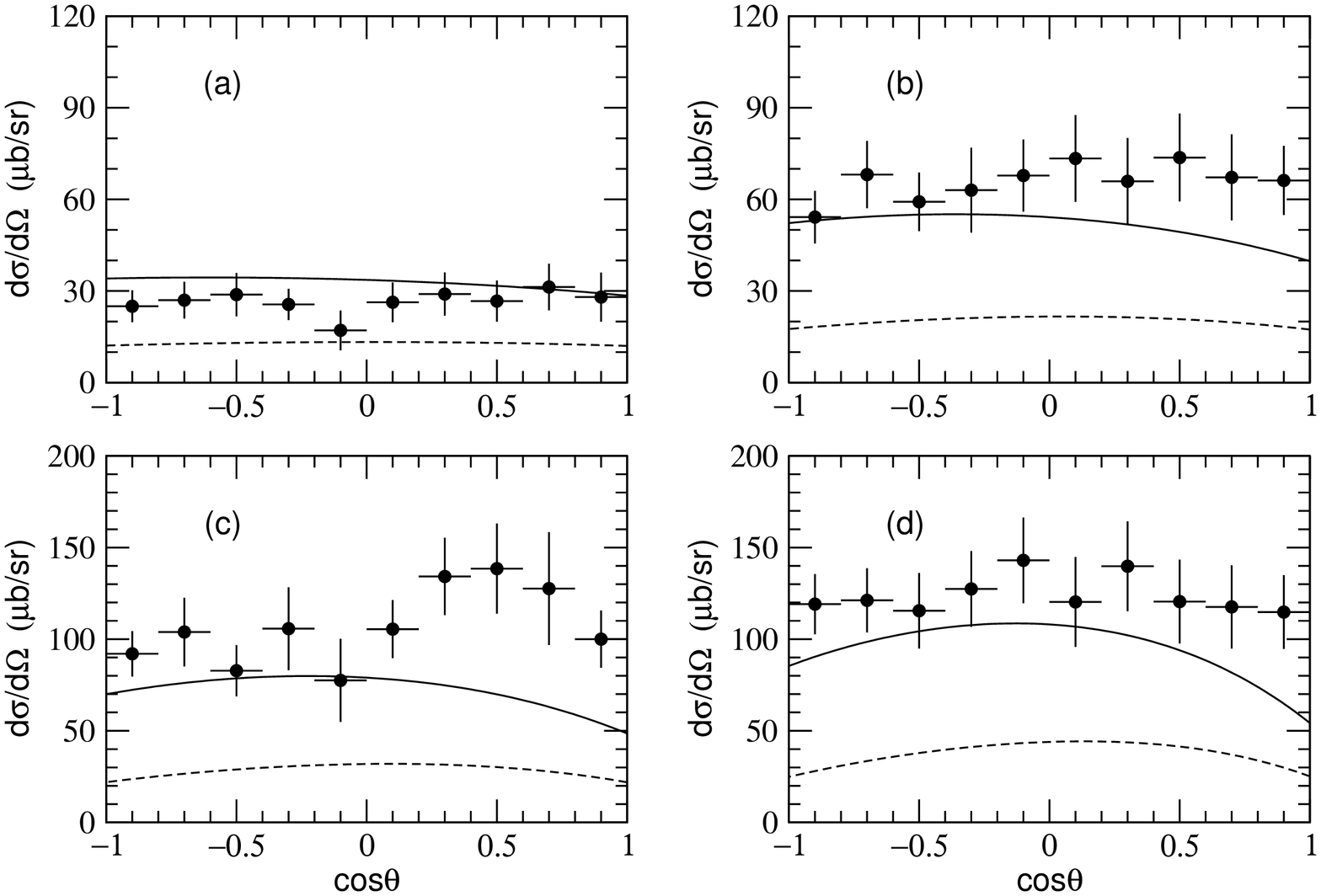, width=9cm, angle=-90}
\caption{Differential cross section for $\pi^- p \to \omega n$ at
$W = $ (a) $1.726$ GeV, (b) $1.734$ GeV, (c) $1.746$ GeV, and (d)
$1.762$ GeV. The solid lines are obtained with $\Lambda_{\omega\rho\pi}
= \Lambda_{\rho NN} = 1.55$ GeV and the dashed lines with
$\Lambda_{\omega\rho\pi} = \Lambda_{\rho NN} = 1.3$ GeV.
The experimental data are from Ref. \cite{KCDG79}.}
\label{fig:pi-om:dcs}
\end{figure}

\begin{figure}
\centering
\epsfig{file=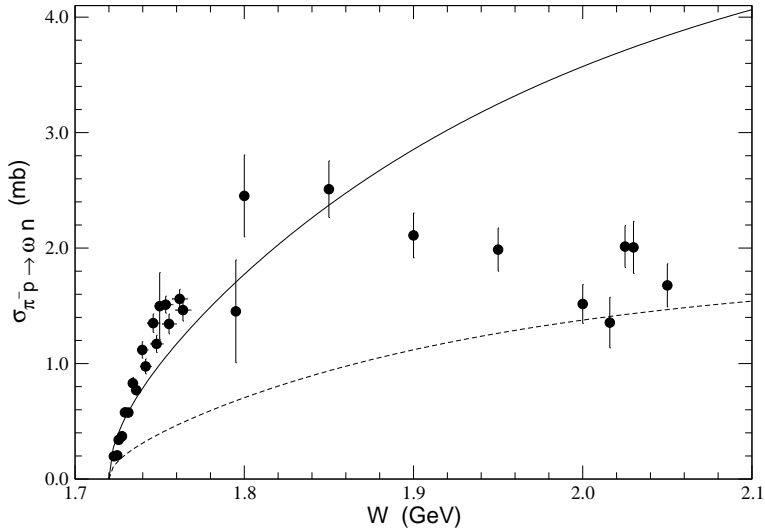, width=9cm, angle=-90}
\caption{Total cross section for $\pi^- p \to \omega n$.
Notations are the same as in Fig.~\ref{fig:pi-om:dcs}.
The experimental data are from Refs. \cite{KCDG79,KBCD76-DADD70}.}
\label{fig:pi-om:tcs}
\end{figure}

In this work, we first consider the one-loop corrections due to the
intermediate $\pi N$ channel in Eq. (\ref{loop-int}).
As discussed in the previous Section, the non-resonant
$B_{\gamma N, \pi N}$ is generated by using the procedure of
Ref. \cite{DGL02} to subtract the resonant amplitudes from the empirical
$\gamma N \rightarrow \pi N$ amplitudes.
Thus our results depend on the employed $\pi N \rightarrow \omega N$
amplitude. 
To proceed, we adjust the form factors of the tree-diagrams of
Fig.~\ref{fig:pi-omega} to fit the $\pi N \rightarrow \omega N$ data.
In addition to the $\rho$ and nucleon exchanges allowed by the Lagrangian
(\ref{lags}), we also consider the exchange of the axial vector 
$b_1(1235)$ meson that was considered to explain the
$\pi N \rightarrow \omega N$ reaction at high energies.
However, we find that its contribution is negligibly small in the
considered energy region.
The details on the $b_1$-exchange amplitude are summarized in Appendix.
Our numerical results show that the $\pi^- p \rightarrow \omega n$
data%
\footnote{For the interpretation of the experimental data of
Ref. \cite{KCDG79}, we follow Ref. \cite{PM01b}.
See also Refs. \cite{TKR01,HK99} for the other interpretation.}
near threshold can be described to some extent by choosing the
following parameters
\begin{equation}
\Lambda_{\omega\rho\pi} = \Lambda_{\rho NN} = 1.55 \mbox{ GeV}, \qquad
\Lambda_{b_1\omega\pi} = \Lambda_{b_1 NN} = 1.4 \mbox{ GeV}, \qquad
\Lambda_N = 0.5 \mbox{ GeV}.
\label{pi-om:1}
\end{equation}
The results are the solid lines in Figs.~\ref{fig:pi-om:dcs} and
\ref{fig:pi-om:tcs}.
In the same figure, we also show the results (dashed curves)
calculated with 
\begin{equation}
\Lambda_{\omega\rho\pi} = \Lambda_{\rho NN} = 1.3 \mbox{ GeV}.
\label{pi-om:2}
\end{equation}
The dashed curve in Fig. \ref{fig:pi-om:tcs} is close to the results from
the coupled-channel $K$-matrix model of Ref. \cite{PM01a} when the
resonance contributions and the coupled-channel effects are neglected.%
\footnote{We are grateful to G. Penner for communications on the results
of Ref. \cite{PM01a}.}
We thus interpret that the model corresponding to the dashed curves
of Fig. \ref{fig:pi-om:tcs} can be used to generate the non-resonant 
$\pi N\rightarrow \omega N$ amplitude for
the calculation according to Eq. (\ref{tGN}) or Eq. (\ref{loop-int}). 


\begin{figure}
\centering
\epsfig{file=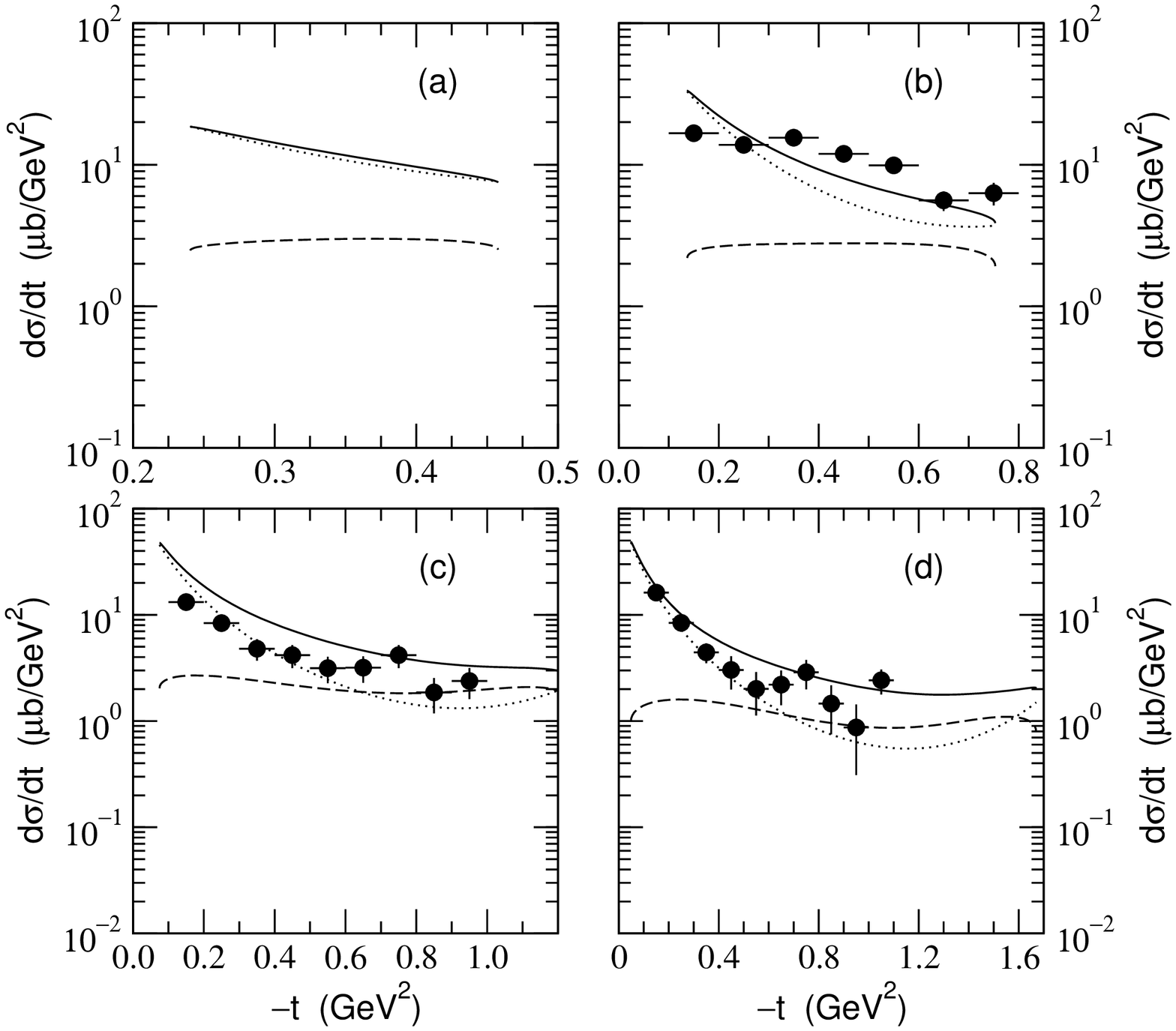, width=9.5cm, angle=-90}
\caption{Differential cross section for $\gamma p \to \omega p$ at
$E_\gamma = $ (a) $1.125$ GeV, (b) $1.23$ GeV, (c) $1.45$ GeV, and (d)
$1.68$ GeV, which corresponds to $W = $ (a) $1.73$ GeV, (b) $1.79$ GeV,
(c) $1.90$ GeV, and (d) $2.01$ GeV, respectively.
The dotted lines are obtained from the tree diagrams and the dashed lines
are from the intermediate $\pi N$ channel.
The solid lines are the sums of the tree diagrams and the intermediate
$\pi N$ channel.
The experimental data are from SAPHIR \cite{Klein96-98}.}
\label{fig:pin:dcs}
\end{figure}

\begin{figure}
\centering
\epsfig{file=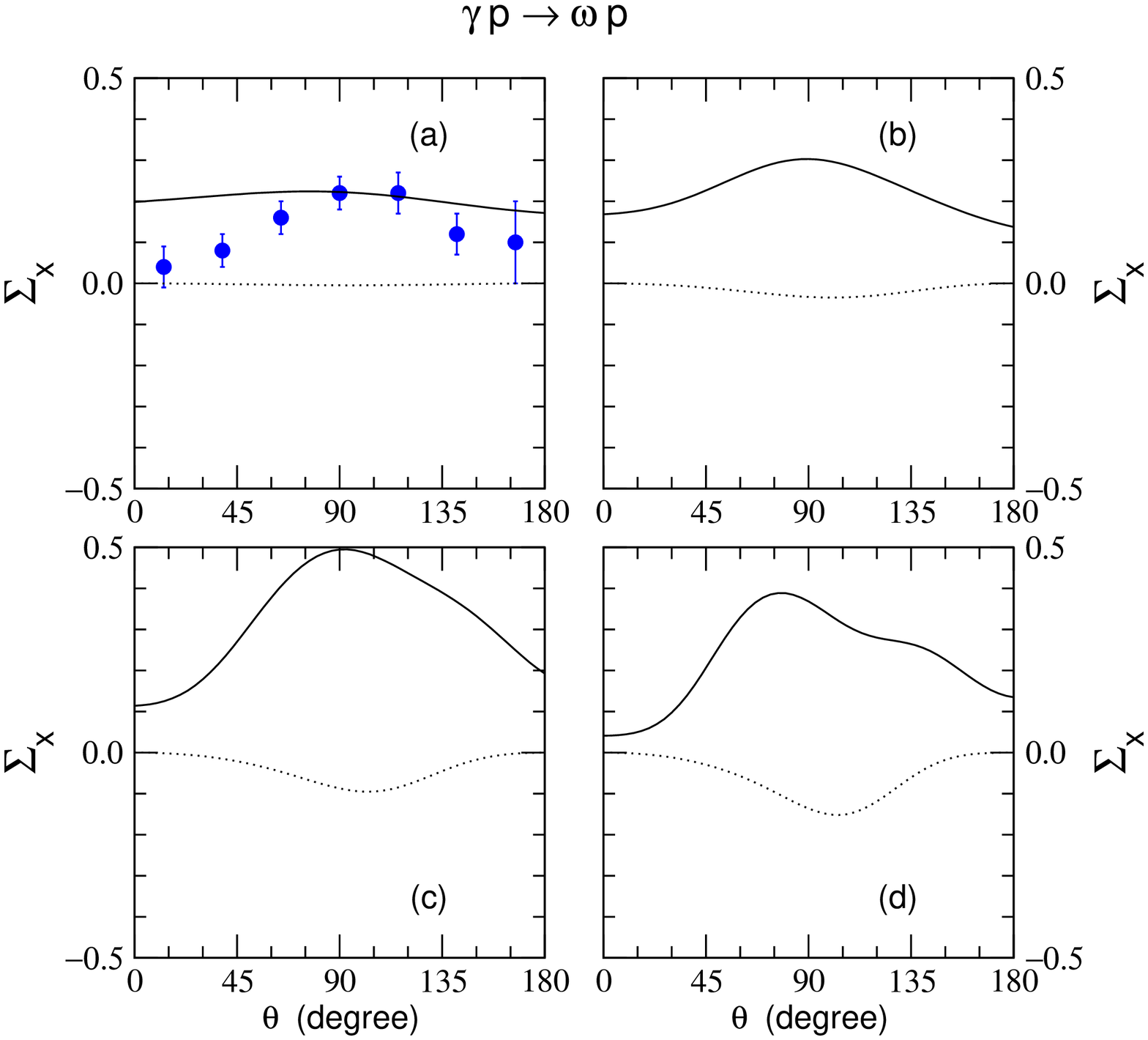, width=9.5cm, angle=-90}
\caption{Single photon asymmetry $\Sigma_x$ for $\gamma p \to \omega p$ at
$E_\gamma = $ (a) $1.125$ GeV, (b) $1.23$ GeV, (c) $1.45$ GeV, and (d)
$1.68$ GeV.
The dotted lines are from the tree diagrams only while the solid lines
include the intermediate $\pi N$ channel.
The experimental data are from Ref. \cite{GRAAL01}}
\label{fig:pol-pin}
\end{figure}

With the non-resonant amplitudes of $\gamma N \rightarrow \pi N$ and
$\pi N \rightarrow \omega N$ transition obtained above, we now use
Eq. (\ref{loop-int}) to compute the one-loop corrections
due to the intermediate $\pi N$ channel.
As shown in Fig.~\ref{fig:pin:dcs}, its magnitudes (dashed lines) are
smaller than those of the tree-diagrams (dotted lines).
However it can have significant effects through its interference with the
tree-diagram amplitude. 
This is evident by comparing the results (solid curves) from the full
calculation and the dotted curves.
The one-loop corrections are even more dramatic in determining the
polarization observables.
An example is shown in Fig. \ref{fig:pol-pin}.
We see that the one-loop corrections
 can change the photon asymmetry
in magnitudes at all angles. 


\begin{figure}
\centering
\epsfig{file=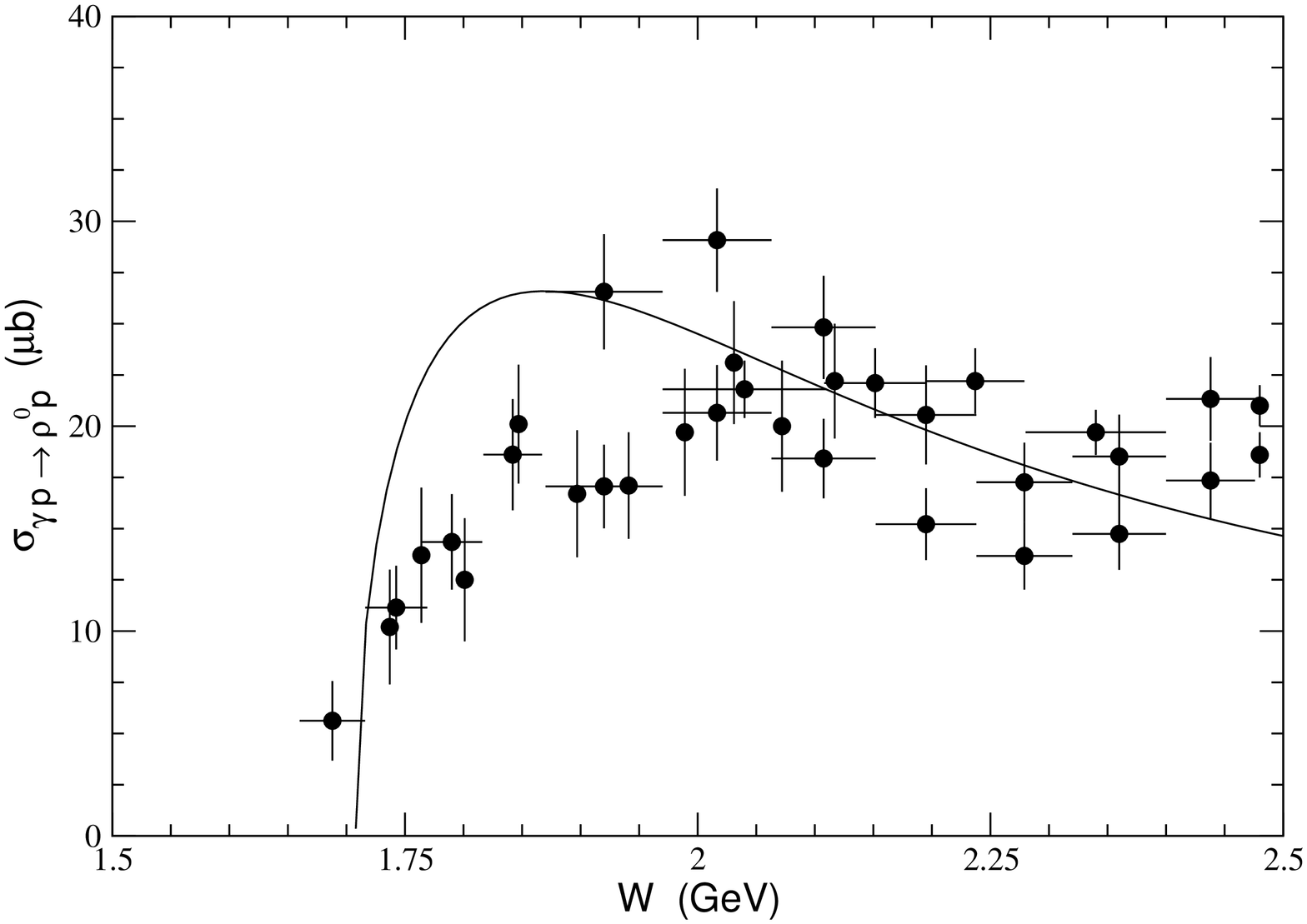, width=8cm, angle=-90}
\caption{Total cross section for $\gamma p \to \rho^0 p$.
The solid line is obtained with the diagrams of Fig.~\ref{fig:gam-rho}.
The experimental data are from Refs.
\cite{SDEJ72,AHHM76,ABBH68,BCGL69-BCGG72-BCEK73,Klein96-98}.}
\label{fig:gam-rho:tcs}
\end{figure}

We now turn to investigating the one-loop corrections due to the
$\rho N$ channel.
{}From the very limited data \cite{AHHM76,BDLM79}, we know that
$\rho^\pm$ photoproduction is much weaker than $\rho^0$ photoproduction.
We therefore only keep $\rho^0 p$ in the loop integration (\ref{loop-int}).
The $\rho\Delta$ channel also is not considered by the same reason.
The $\gamma p \rightarrow \rho^0p$ amplitude is generated from
the tree diagrams in Fig. \ref{fig:gam-rho} and the
$\rho^0 p \rightarrow \omega p$ amplitude from the tree diagrams in
Fig. \ref{fig:rho-omega}.
We note that the tree-diagrams in Fig. \ref{fig:gam-om} and
Fig. \ref{fig:rho-omega} are related in the vector dominance model, except
that  the Pomeron and the $\eta$ exchanges are not allowed in
$\rho^0 p \rightarrow \omega p$ transition because of their quantum numbers.

\begin{figure}
\centering
\epsfig{file=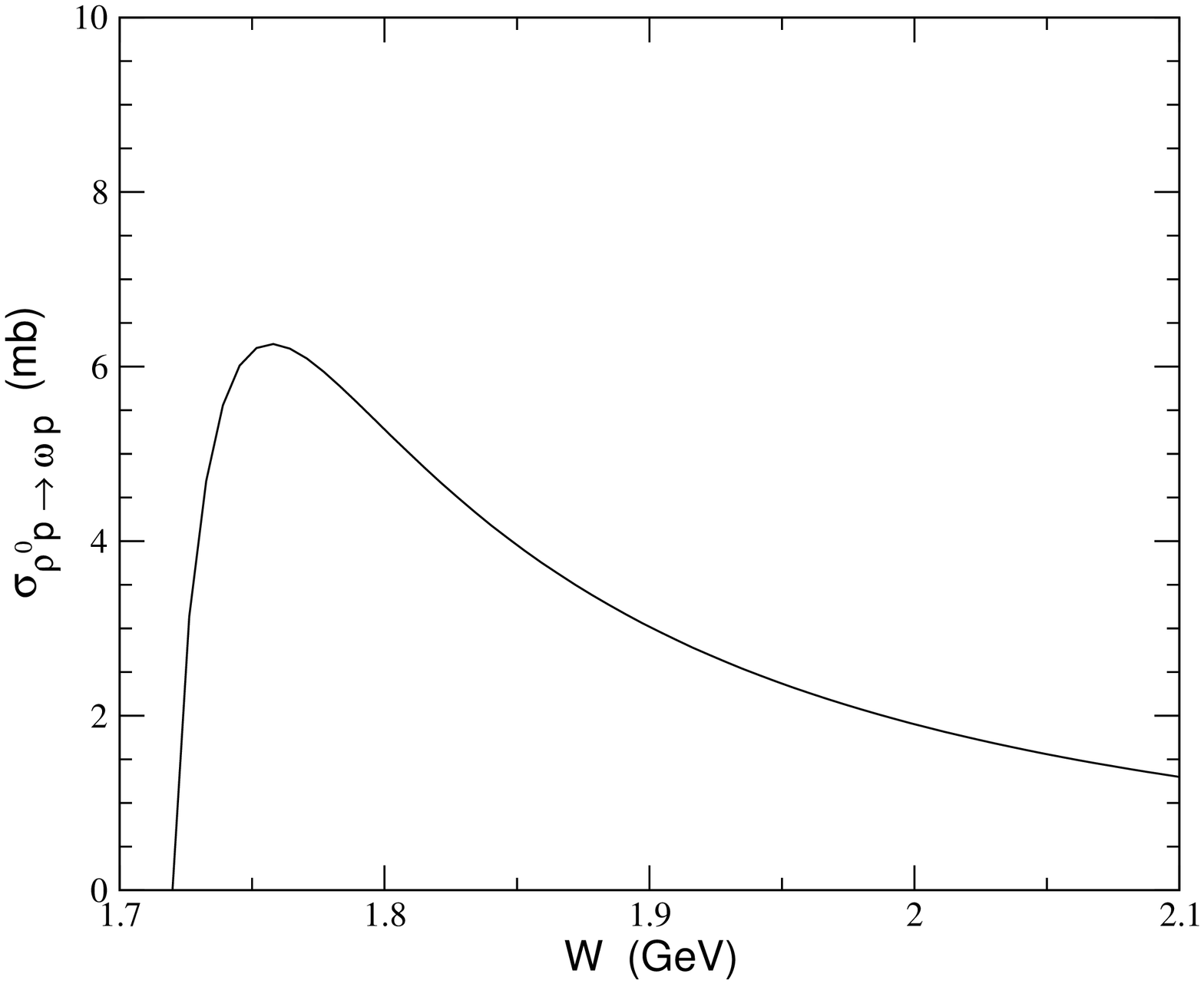, width=8cm, angle=-90}
\caption{Total cross section for $\rho^0 p \to \omega p$.
The solid line is the result of the diagrams shown in
Fig.~\ref{fig:rho-omega}.}
\label{fig:rho-omega:tcs}
\end{figure}

We find that the constructed $\gamma p \rightarrow \rho^0 p$ amplitude
(Fig.~\ref{fig:gam-rho}) can reproduce the total cross section data,
if we use the following cutoff parameters (in unit of GeV) \cite{OTL00}
\begin{equation}
\Lambda_{\pi NN} = 0.6, \qquad
\Lambda_{\rho\pi\gamma} = 0.77, \qquad
\Lambda_{\sigma NN} = 1.0, \qquad
\Lambda_{\sigma\rho\gamma} = 0.9, \qquad
\Lambda_N = 0.5.
\end{equation}
Our results are shown in Fig.~\ref{fig:gam-rho:tcs}.
On the other hand, there is no data to constrain our model for
$\rho^0 p \to \omega p$ (Fig.~\ref{fig:rho-omega}).
Motivated by vector dominance model, we therefore
calculate these tree
diagrams using the same form factors, given in Eq. (\ref{cutoff:gam-om}),
of Fig.~\ref{fig:gam-om} for $\omega$ photoproduction.
The predicted total cross sections of $\rho^0 p \rightarrow \omega p$
reaction are shown in Fig.~\ref{fig:rho-omega:tcs}.
We find that its magnitude at the peak is a factor of about $3$ larger
than that in Fig. \ref{fig:pi-om:tcs} for the $\pi^- p  \to \omega n$
reaction.
This assumption may lead to an unrealistic estimation of the
one-loop corrections due to $\rho N$ channel.
Another uncertainty in the calculation of Eq. (\ref{loop-int}) with
$\rho N$ intermediate state is that the correct input to the loop
integration (\ref{loop-int}) is the non-resonant amplitude, not the full
amplitudes constructed above.
But there is no experimental information we can use here to extract the
non-resonant part from the full amplitude.
For these reasons, we perform the $\rho N$
loop integration (\ref{loop-int})
using the constructed full amplitudes of both the
$\gamma p \rightarrow \rho^0 p$ and $\rho^0 p \rightarrow \omega p$
transitions. Therefore, our results for the $\rho N$ loop can only be
 considered as an upper bound.


\begin{figure}
\centering
\epsfig{file=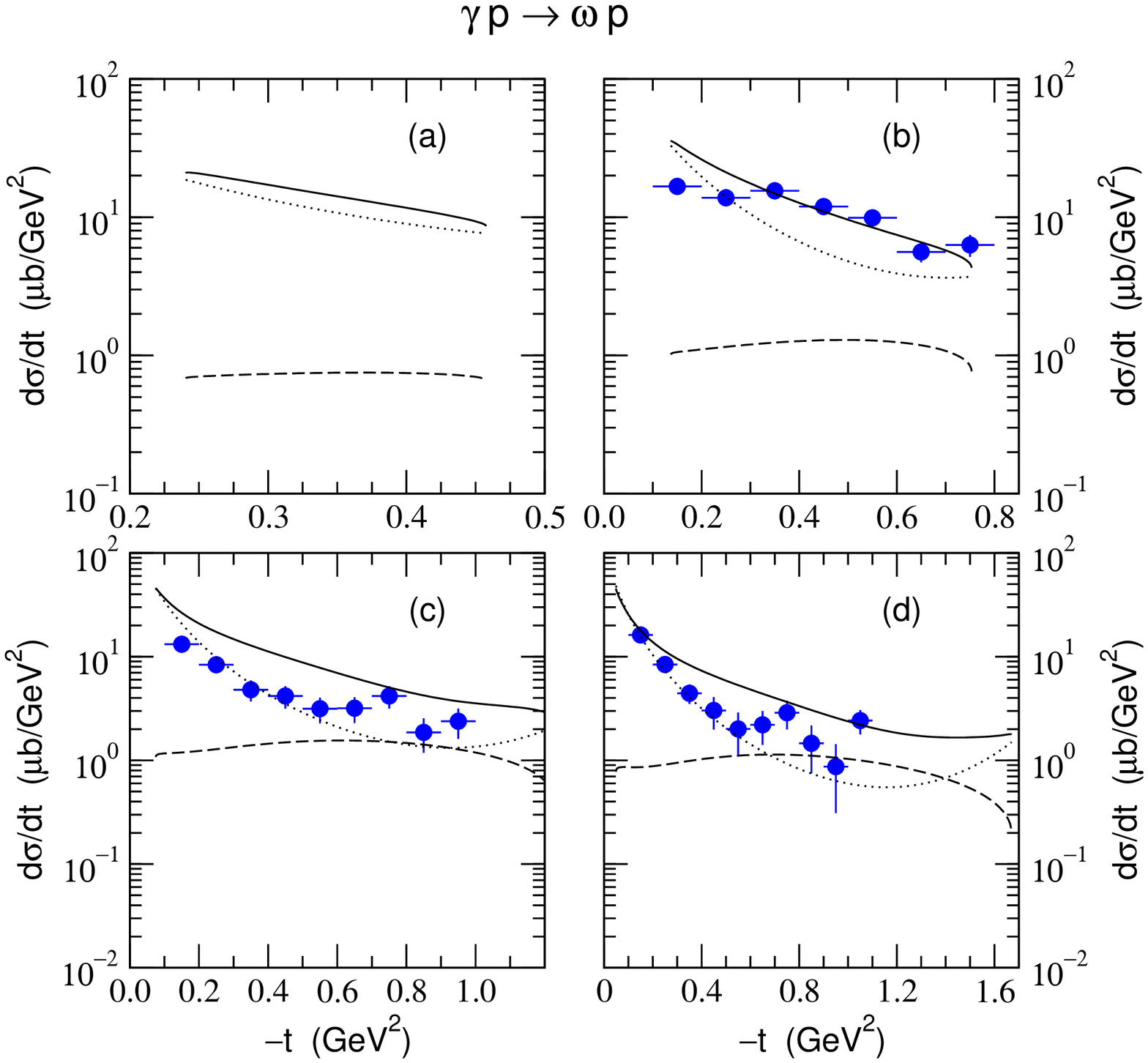, width=9.5cm, angle=-90}
\caption{Differential cross section for $\gamma p \to \omega p$ at
$E_\gamma = $ (a) $1.125$ GeV, (b) $1.23$ GeV, (c) $1.45$ GeV, and (d)
$1.68$ GeV.
The dotted lines are from the tree diagrams and the dashed lines from
the intermediate $\rho N$ channel.
The solid lines are the results including the tree diagrams and the
intermediate $\pi N$ and $\rho N$ channels.
The experimental data are from SAPHIR \cite{Klein96-98}.}
\label{fig:rhoN:dcs}
\end{figure}

\begin{figure}
\centering
\epsfig{file=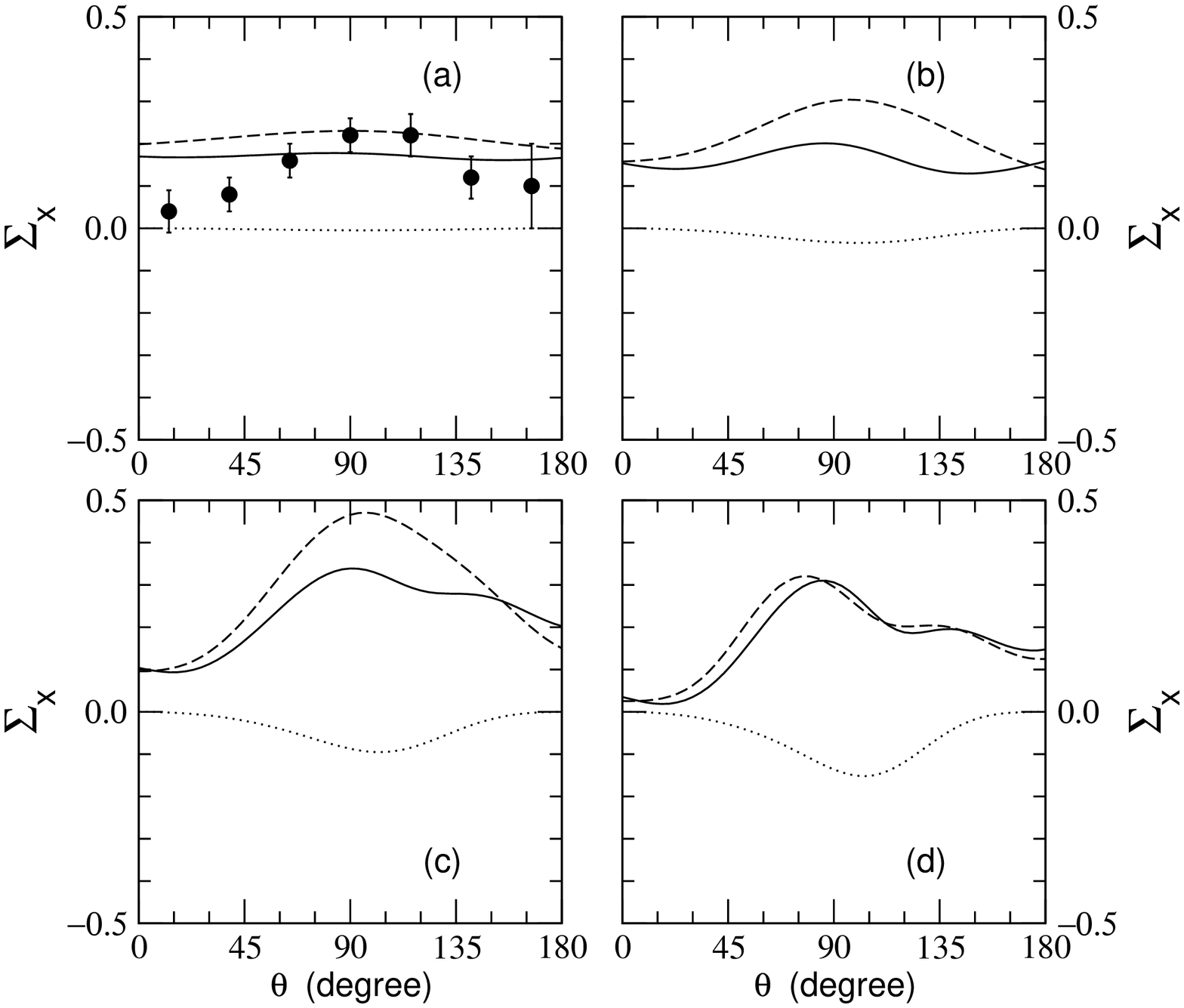, width=9.5cm, angle=-90}
\caption{Single photon asymmetry $\Sigma_x$ for $\gamma p \to \omega p$
at
$E_\gamma = $ (a) $1.125$ GeV, (b) $1.23$ GeV, (c) $1.45$ GeV, and (d)
$1.68$ GeV.
The dotted lines are from the tree diagrams only and the dashed lines
include the intermediate $\pi N$ channel.
The full calculations including the tree diagrams and the intermediate
$\pi N$ and the $\rho N$ channels are given by the solid lines.
The experimental data are from Ref. \cite{GRAAL01}}
\label{fig:pol-all}
\end{figure}

The calculated 
 one-loop corrections due to the $\rho N$ channel are shown in
Fig.~\ref{fig:rhoN:dcs}. 
Comparison with the results given in Fig.~\ref{fig:pin:dcs} shows that
the effects of the $\rho N$ channel are as large as or even bigger than
those of the intermediate $\pi N$ channel.
Thus the full calculation (solid lines) including both $\pi N$ and
$\rho N$ channels gives large corrections to the tree-diagram results
(dotted lines). 
The corresponding coupled-channel effects on photon asymmetry 
are shown in Fig. \ref{fig:pol-all}.
Again, we see that the polarization effects are sensitive to the
one-loop corrections.


\section{Conclusions}

As a step toward developing a coupled-channel 
model of vector meson photoproductions, the 
one-loop corrections to $\omega$ photoproduction have been investigated.
The calculations
have been performed by assuming that all relevant non-resonant 
amplitudes can be calculated from tree-diagrams
 of effective Lagrangians. Our calculation of the
one-loop corrections due to the intermediate $\pi N$ channel
is rather well constrained by the data of $\gamma N \rightarrow \pi N$ and
$\pi N \rightarrow \omega N$ reactions. On the other hand, 
our treatment of $\rho N$ channel involves some uncertainties, mainly
due to the lack of enough experimental inputs such as the data of
$\rho N \rightarrow \omega N$ reaction. Therefore,
our results for $\rho N$ channel can only be regarded as an upper bound.

As discussed in section II, the one-loop corrections are just the
leading terms of a perturbative expansion of a full coupled-channel
 model.  Thus the results presented in this paper can only be taken 
as a qualitative
indication of the importance of channel coupling effects. 
We have shown  that the one-loop corrections due to intermediate
$\pi N$ and $\rho N$ channels  are comparable to those of nucleon 
resonance contributions investigated in Ref. \cite{OTL01}. The
results from this rather exploratory investigation
suggest strongly that the coupled-channel
 effects should be carefully taken into account in extracting 
the resonance parameters from the experimental data, in particular the
data of polarization observables.


\acknowledgments

Y.O. is grateful to the Physics Division of Argonne National Laboratory
for the hospitality.
This work was supported in part by the Brain Korea 21 project of Korean
Ministry of Education, the International Collaboration Program of
KOSEF under Grant No. 20006-111-01-2, and U.S. DOE Nuclear Physics
Division Contract No. W-31-109-ENG-38.


\appendix*

\section{The axial $\bm{\lowercase{b}_1(1235)}$ meson exchange in
$\bm{\pi N \to \omega N}$}

In this Appendix, we discuss the axial vector $b_1(1235)$ meson
exchange in $\pi N \to \omega N$ reaction.
Recently, this reaction has been studied by effective Lagrangian
method and unitary coupled channel models focusing on the role of
the nucleon resonances \cite{LWF99,LWF01,TKR01,PM01,PM01a}.
In the early investigations in 1960's and 1970's this reaction was
studied in some detail mostly based on Regge theory and absorption models
and at higher energies \cite{JDGK65,Barm66-68,HKK68,HK71,IM74}.
Based on the Regge theory, the $b_1$ trajectory exchange has been
discussed as the secondary exchange process in $\pi N \to \omega N$
in addition to the major $\rho$-trajectory exchange.
The main motivation for the secondary exchange was to account for the
experimentally observed non-vanishing vector meson density matrix
$\rho_{00}$ that is expected to vanish if the natural-parity
$\rho$-trajectory exchange dominates.
The $b_1(1235)$ meson has quantum numbers $I^G(J^{PC}) = 1^+ (1^{+-})$
with mass $M_{b_1}=1230$ MeV and width $\Gamma_{b_1}=142 \pm 9$ MeV,
and it mostly decays into the $\omega\pi$ channel \cite{PDG00}.
Thus its exchange can contribute to $\pi N \to \omega N$ as an unnatural
parity exchange.
In this work we consider the one-$b_1$-exchange process (not the
exchange of $b_1$ trajectory) in $\pi N \to \omega N$.

The general form of the $b_1 \omega \pi$ interaction can be written as
\cite{BD63}
\begin{equation}
\mathcal{M}_{b_1 \omega \pi} = -i M_{b_1} \varepsilon^*_\mu(\omega) \left[ f
g^{\mu\nu} + \frac{h}{M_\omega M_{b_1}} q^\nu k^\mu \right]
\varepsilon_\nu (b_1),
\label{b-om-pi}
\end{equation}
where $k$ and $q$ are the momenta of the $b_1$ and $\omega$,
respectively, and $\varepsilon_\mu(b_1)$ and $\varepsilon_\mu(\omega)$
are their polarization vectors.
Then the total decay width reads
\begin{equation}
\Gamma_{b_1 \to \omega\pi} = \frac{|{\bf q}|}{24\pi} \left\{ 2 f^2 +
\frac{1}{M_\omega^4} \left( E_\omega M_\omega f + |{\bf q}|^2 h
\right)^2 \right\},
\end{equation}
where $E_\omega$ is the $\omega$ meson energy in the $b_1$ rest frame.
The unknown coupling constants $f$ and $h$ can then be determined by the
decay width and the $D/S$ amplitude ratio in the decay of $b_1 \to
\omega \pi$, where Eq. (\ref{b-om-pi}) gives \cite{IMR89}
\begin{equation}
f^D/f^S = - \frac{\sqrt2 \left\{ M_\omega (E_\omega - M_\omega) f +
|{\bf q}|^2 h \right\}}{ M_\omega (E_\omega + 2 M_\omega) f + |{\bf
q}|^2 h},
\end{equation}
which is defined from
\begin{equation}
\langle \omega({\bf q},m_\omega) \pi(-{\bf q}) | H_{\rm int} | b_1 ({\bf
0}, m_b) \rangle
= i f^S \delta_{m_\omega m_b} Y_{00}(\Omega_q)
+ i f^D \sum_{m_\ell} \langle 2\, m_\ell \, 1 \, m_\omega | 1 \, m_b \rangle
Y_{2 m_\ell}(\Omega_q),
\end{equation}
where $Y_{\ell m}(\Omega)$ and $\langle j_1\, m_1\, j_2\, m_2 \mid j\, m
\rangle$ are the spherical harmonics and Clebsch-Gordan coefficients, and
$m_\omega$ ($m_b$) is the spin projection along the $z$ axis for the
$\omega$ ($b_1$) meson.
Using the PDG \cite{PDG00} values for $\Gamma_{b_1 \to \omega \pi}$ and
$f^D/f^S$, i.e., $\Gamma_{b_1\to \omega\pi} = 142 \pm 9$ MeV and
$f^D/f^S = 0.29 \pm 0.04$, we obtain
\begin{equation}
f \approx 3.71, \qquad h \approx -11.38,
\end{equation}
which gives $h/f \approx -3.1$.
This should be compared with the value $h/f = +9.0$ used by
Ref. \cite{Barm66-68} to fit the high energy data of
$\pi N \to \omega N$ together with the $\rho$-trajectory exchange.

The $b_1$-nucleon coupling can be written as
\begin{equation}
\mathcal{M}_{b_1 NN} = \frac{ig_{b_1NN}^{}}{2M_N} \bar{\psi} \sigma^{\mu\nu}
\gamma_5 q^\nu \bm{\tau} \cdot \bm{b}_\mu \psi,
\end{equation}
due to the $G$ parity of the $b_1$, where $b_\mu$ is the $b_1$ meson
field and $\psi$ is the nucleon. The momentum of the $b_1$ meson is
denoted by $q_\mu$.
The coupling constant $g_{b_1 NN}^{}$ is related to the nucleon tensor
charge and has been recently estimated by making use of the axial vector
dominance and SU(6) $\times$ O(3) spin-flavor symmetry in
Ref. \cite{GGo01} in a similar way to Ref. \cite{BF96}.
(See also Ref. \cite{KMOV00}.)
The result reads
\begin{equation}
g_{b_1NN}^{} = \frac{5}{3\sqrt2} g_{a_1NN}^{}.
\end{equation}
Using $g_{a_1NN}^{} \approx 7.49$, one finally obtains
$g_{b_1NN}^{} \approx 8.83$ \cite{GGo01}.

Thus the production amplitude for the reaction of $\pi N \to \omega N$
is obtained as
\begin{eqnarray}
\mathcal{M} &=&  \frac{M_{b_1} g_{b_1NN}^{} C_I}{2M_N [ (k-q)^2 - M_{b_1}^2]}
\varepsilon^*_\mu(\omega) \left\{ f g^{\mu\nu} + \frac{h}{M_\omega
M_{b_1}} q^\nu (k-q)^\mu \right\}
\nonumber \\ && \mbox{} \times
\left\{ g_{\nu\alpha} - \frac{(k-q)_\nu (k-q)_\alpha}{M_{b_1}^2}
\right\} \bar{u}(p') \gamma_5 \sigma^{\alpha\beta} (k-q)_\beta u(p),
\end{eqnarray}
where $u(p)$ is the Dirac spinor of the nucleon with momentum $p$.
The isospin factor $C_I$ is
\begin{equation}
C_I = \left\{ \begin{array}{ll} \sqrt2 \qquad & \mbox{ for } \pi^- p \to
\omega n,\  \pi^+ n \to \omega p \\
+1 & \mbox{ for } \pi^0 p \to \omega p \\
-1 & \mbox{ for } \pi^0 n \to \omega n \end{array} \right.
\label{isofactor}
\end{equation}


\begin{figure}
\centering
\epsfig{file=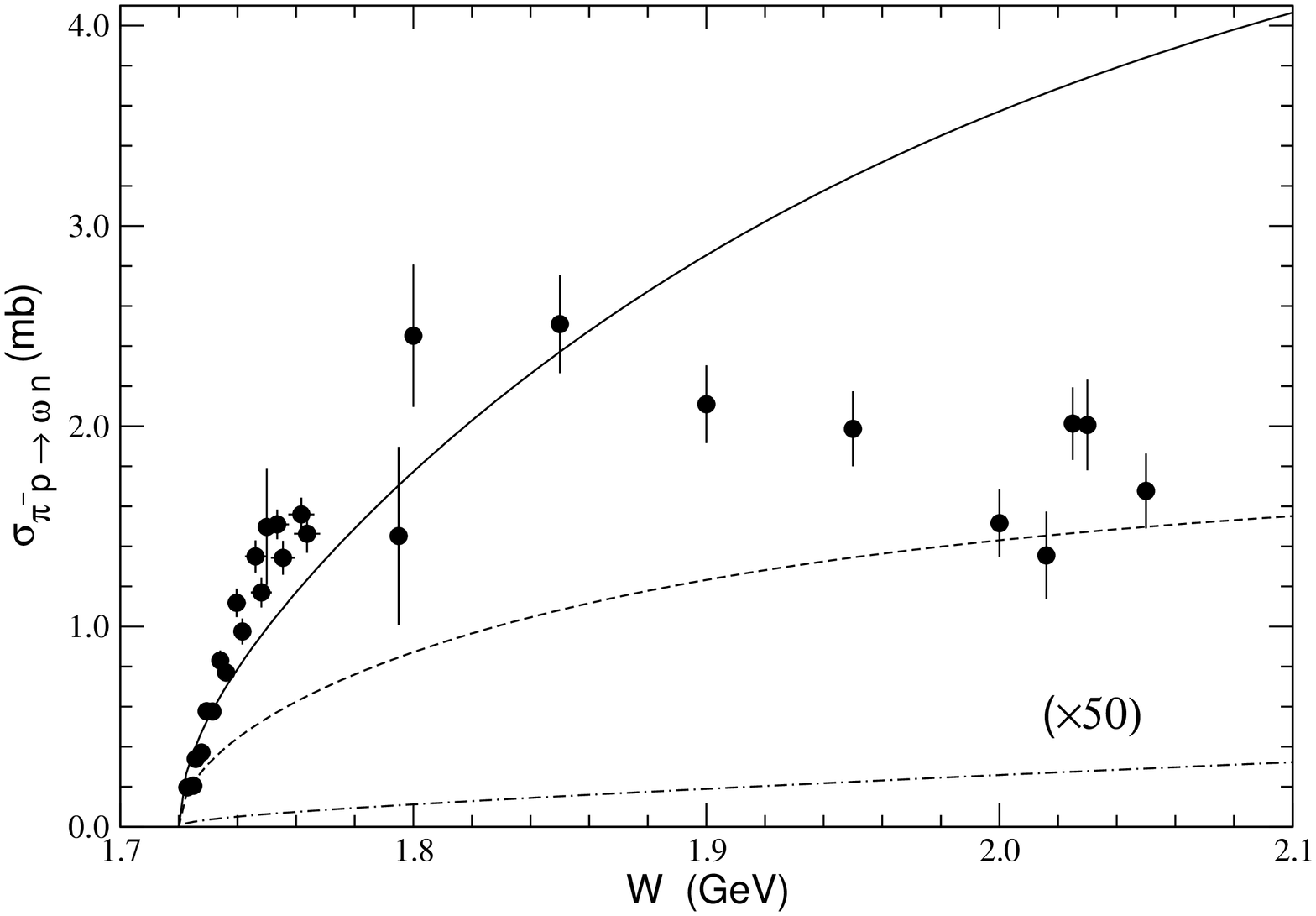, width=9cm, angle=-90}
\caption{Total cross section for $\pi^- p \to \omega n$.
Solid and dashed lines are the same as in Fig.~\ref{fig:pi-om:dcs} while the
dot-dashed line is from the $b_1(1235)$ exchange process (multiplied by
$50$).
The experimental data are from Refs. \cite{KCDG79,KBCD76-DADD70}.}
\label{fig:pi-om-b1:tcs}
\end{figure}

Given in Fig.~\ref{fig:pi-om-b1:tcs} are the total cross sections for
$\pi^- p \to \omega n$.
The solid and dashed lines are obtained with the $\rho$ exchange and the
nucleon pole terms with the cutoff parameters (\ref{pi-om:1}) and
(\ref{pi-om:2}), respectively.
The total cross section due to the $b_1$ exchange is given by the
dot-dashed line.
For the form factor, we use the form of Eq.~(\ref{Ft}) with
$\Lambda_{b_1\omega\pi} = \Lambda_{b_1 NN} = 1.4$ GeV.
Since its contribution is suppressed by $\rho$ and nucleon exchange
contributions, the $b_1$ exchange is magnified in
Fig.~\ref{fig:pi-om-b1:tcs} by a factor of $50$.
This conclusion does not sensitively depend on the cutoff parameters
$\Lambda_{b_1\omega\pi}$ and $\Lambda_{b_1 NN}$, when they are larger
than the exchanged meson mass $M_{b_1}$.


\newpage


\end{document}